\documentclass{iopart}
\usepackage{iopams}
\usepackage{amssymb}
\usepackage{amscd}
\usepackage{bm}
\usepackage{graphicx}
\begin{document}

\title{Gravitational waves carrying orbital angular momentum}
\author{Iwo Bialynicki-Birula}
\address{Center for Theoretical Physics, Polish Academy of Sciences,\\Al. Lotnik\'ow 32/46, 02-668 Warsaw, Poland}
\ead{birula@cft.edu.pl}
\author{Zofia Bialynicka-Birula}
\address{Institute of Physics, Polish Academy of Sciences\\
Aleja Lotnik\'ow 32/46, 02-668 Warsaw, Poland}

\begin{abstract}
Spinorial formalism is used to map every electromagnetic wave into the gravitational wave (within the linearized gravity). In this way we can obtain the gravitational counterparts of Bessel, Laguerre-Gauss, and other light beams carrying orbital angular momentum.
\end{abstract}
\submitto{New Journal of Physics}

\section{Introduction}

Beams of electromagnetic radiation endowed with orbital angular momentum (OAM) have become objects of intense theoretical and experimental studies \cite{oam}. The aim of this work is to construct explicitly new solutions of the field equations in the {\em linearized gravity} which describe the gravitational waves carrying orbital angular momentum. More precisely, these solutions have a specific value of the angular momentum; they are eigenfunctions (labeled by the integer quantum number $m$) of the component of the angular momentum operator in the direction of propagation. We shall restrict ourselves to the propagation of waves in empty space. The problem of the generation of such waves would require a separate study.

The gravitational waves will by constructed using the correspondence between electromagnetism and gravitation. In a recent publication in this journal Stephen Barnett \cite{sb} addressed the problem of such a correspondence. He argued that the counterpart of the electromagnetic field tensor is an object constructed from the first derivatives of the metric tensor.

In the present paper we use a different electromagnetic-gravitational correspondence that has been used in previous studies of this problem in general relativity \cite{mb,owen,kt,kt1,kt2} and also in the construction of gravitational Hopfions---knotted gravitational waves \cite{tsb,twsb,stwdb}. This paradigm is built starting from the correspondence between the wave equations for photons and gravitons. This leads to the unavoidable conclusion that the counterpart of the electromagnetic field tensor is the gravitational Riemann curvature tensor rather than the first derivatives of the metric tensor as proposed in \cite{sb}.

We shall fully employ the spinorial formalism \cite{pr,cors} which is most useful in this context because it enables us to establish a {\em one-to-one} correspondence between the electromagnetic waves and the gravitational waves. Using this formalism we can obtain in a straightforward way the counterparts of all important electromagnetic waves carrying orbital angular momentum: Bessel beams, Laguerre-Gauss beams, exponential beams, multipole fields, etc. The electromagnetic and gravitational waves will be built with the use of a technique originated by Hertz and then improved by many others. Following John Stewart \cite{js} we shall call it the HBDWP method and we will use its most refined version due to Roger Penrose \cite{rp}. The characteristic feature of the HBDWP method is the presence of the generating function: a solution of the scalar d'Alembert equation. The complete wave is obtained by simply taking various combinations of the derivatives of the generating function. The one-to-one correspondence between electromagnetic and gravitational waves is established by using {\em the same} generating function in both cases. We shall illustrate our procedure with the construction of gravitational Bessel beams.

\section{Spinorial formalism}

Wave fields in spacetime in the massless case can be described by symmetric spinors whose rank is equal to twice the absolute value of helicity; for the electromagnetic field the rank is two and for the gravitational field the rank is four. These spinors obey in empty space the wave equations of the same form for every helicity \cite{pr},
\begin{eqnarray}\label{weq}
g^{\mu{\dot A}A}\partial_\mu\phi_{AB\dots L}(\bm r,t)=0,
\end{eqnarray}
where the indices take on the values 0 and 1, dotted indices refer to complex conjugate transformation properties, and repeated indices imply summation. The spin-vectors $g^{\mu{\dot A}B}$ and $g^{\mu}_{\;A\dot{B}}$ are the Pauli matrices complemented by the unit matrix,
\numparts
\begin{eqnarray}\label{gu}
g^{0{\dot A}B}=I,\quad g^{1{\dot A}B}=\sigma_x,\;\quad g^{2{\dot A}B}\!=\sigma_y,\;\quad g^{3{\dot A}B}=\sigma_z,\\
g^{0}_{\;A\dot{B}}=I,\quad g^{1}_{\;A\dot{B}}=-\sigma_x,\quad g^{2}_{\;A\dot{B}}=-\sigma_y,\quad g^{3}_{\;A\dot{B}}=-\sigma_z.
\end{eqnarray}
\endnumparts
Spinor indices are raised and lowered with the use of the metric spinors $\epsilon_{AB}$ and $\epsilon^{AB}$,
\begin{eqnarray}\label{lr}
\psi^A=\epsilon^{AB}\psi_B,\quad \psi_B=\psi^A\epsilon_{AB},\\
\epsilon_{AB}=\left[\begin{array}{cc}
0&1\\-1&0
\end{array}\right]=\epsilon^{AB}
\end{eqnarray}
The same rules hold for dotted indices.

We shall also need the spin-tensor $S_{\mu\nu}^{\;\;AB}$ constructed from the spin-vectors:
\begin{eqnarray}\label{ss0}
S_{\mu\nu}^{\;\;AB}=\frac{1}{2}\epsilon^{AC}\left(
g_{\mu C\dot{D}}g_{\nu}^{\;\dot{D}B}-g_{\nu C\dot{D}}g_{\mu}^{\;\dot {D}B}\right)
\end{eqnarray}
or more explicitly
\begin{eqnarray}\label{ss}
\fl\qquad\{S_{01},S_{02},S_{03}\}=-\rmi\sigma_y\{\sigma_x,\sigma_y,\sigma_z\},\quad \{S_{23},S_{31},S_{12}\}=\rmi\{S_{01},S_{02},S_{03}\}.
\end{eqnarray}

It has been shown by Penrose \cite{rp} that every solution of the wave equation (\ref{weq}) can be written in the form
\begin{eqnarray}\label{sol}
\phi_{AB\dots L}(\bm r,t)=g^{\mu}_{\;A\dot{A}}\sigma^{\dot A}g^{\nu}_{\;B\dot{B}}\sigma^{\dot B}\dots g^{\lambda}_{\;L\dot{L}}\sigma^{\dot L}\partial_\mu\partial_\nu\dots\partial_\lambda\,\chi(\bm r,t),
\end{eqnarray}
where $\chi(\bm r,t)$ is the generating function: a complex solution of the scalar d'Alembert equation and $\sigma^{\dot A}$ is some constant spinor. To prove that (\ref{sol}) satisfies (\ref{weq}) is an easy task but to prove that every solution has this form is much harder \cite{rp}. Different choices of $\sigma^{\dot A}$ give solutions that are related by a Lorentz transformation. We will make the simplest choice: $\sigma^{\dot A}=\{1,0\}$. In this case
\begin{eqnarray}\label{sol1}
g^{\mu}_{\;A\dot{A}}\sigma^{\dot A}\partial_\mu=\{\partial_t-\partial_z,-\partial_x-\rmi\partial_y\},
\end{eqnarray}
and we obtain for the component of the spinor $\phi_{AB\dots L}$ that has $l$ times index 0 and $n$ times index 1 the formula:
\begin{eqnarray}\label{sol2}
\phi_{00\dots011\dots 1}
=\left(\partial_t-\partial_z\right)^l\left(-\partial_x-\rmi\partial_y\right)^n\chi(\bm r,t).
\end{eqnarray}
Ordering of the indices 0 and 1 does not matter because $\phi_{AB\dots L}$ is fully symmetric. The final expression for the solution of the wave equation will be particularly simple when all the multiple derivatives are easily evaluated. This is precisely the case for the Bessel beams.

\section{The electromagnetic field}

In the spinorial formalism the electromagnetic field tensor is represented by a complex symmetric second rank spinor $\phi_{AB}(\bm r,t)$. This spinor is directly related to the self-dual complex electromagnetic field tensor $F_{\mu\nu}$,
\begin{eqnarray}\label{sd}
F_{\mu\nu}(\bm r,t)
=f_{\mu\nu}+\frac{\rmi}{2}\epsilon_{\mu\nu\alpha\beta}f^{\alpha\beta},
\end{eqnarray}
where $\epsilon_{\mu\nu\alpha\beta}$ is the completely antisymmetric tensor ($\epsilon_{0123}=-1$).
The relation between $F_{\mu\nu}$ and $\phi_{AB}$ reads:
\begin{eqnarray}\label{sd1}
F_{\mu\nu}(\bm r,t)=S_{\mu\nu}^{\;\;AB}\phi_{AB}(\bm r,t),
\end{eqnarray}
Due to self-duality of $F_{\mu\nu}$, only half of this tensor, say $F_{0k}$, carries information; the other half can be easily recovered, $F_{ij}=\rmi\epsilon_{ijk}F_{0k}$. We shall use the Riemann-Silberstein vector ${\bm F}=\{F_{01},F_{02},F_{03}\}$ as a convenient carrier of complete information about the electromagnetic field \cite{pwf,bb1}. It follows from (\ref{sd}) that
\begin{eqnarray}\label{rs}
{\bm F}={\bm E}+\rmi c{\bm B}.
\end{eqnarray}

The three-dimensional complex Riemann-Silberstein vector is a relativistic object owing to the one-to one correspondence between the components of $f_{\mu\nu}$ and $\bm F$. The transformation properties of $f_{\mu\nu}$ under the full Lorentz group induce the following transformation properties of $\bm F$ \cite{pwf}:
\begin{eqnarray}\label{oij}
F'_i(\bm r',t')=O_{ij}(\Lambda)F_j(\bm r,t).
\end{eqnarray}
For three-dimensional rotations $O_{ij}(\Lambda)$ is the standard orthogonal rotation matrix, while for special Lorentz transformation $O_{ij}(\Lambda)$ is the following complex orthogonal matrix:
\begin{eqnarray}\label{rstr}
O_{ij}(\Lambda)=\gamma\left(\delta_{ij}-i\epsilon_{ijk}v_k/c
-\frac{\gamma}{1+\gamma}v_iv_j/c^2\right),
\end{eqnarray}
where $\gamma=1/\sqrt{1-{\bm v}^2/c^2}$.

The components of the Riemann-Silberstein vector are related to the components of the spinor $\phi_{AB}$ as follows:
\begin{eqnarray}\label{phi}
F_x=\phi_{11}-\phi_{00},\quad F_y=-\rmi(\phi_{11}+\phi_{00}),\quad F_z=2\phi_{01}.
\end{eqnarray}
Once we determine the components of $\phi_{AB}$, the electric and magnetic field vectors are obtained by taking the real and imaginary parts of (\ref{phi}).

\section{The gravitational field}

Following the electromagnetic example, we shall now extend the Riemann curvature tensor\footnote{We use the term Riemann tensor even though in empty space it is rather the Weyl tensor since the Ricci tensor $R_{\mu\nu}$ vanishes.} to its self-dual counterpart,
\begin{eqnarray}\label{sdg}
G_{\mu\nu\lambda\rho}=R_{\mu\nu\lambda\rho}
+\frac{\rmi}{2}\epsilon_{\mu\nu\alpha\beta}R_{\lambda\rho}^{\;\;\;\alpha\beta}.
\end{eqnarray}
In the spinorial formalism this tensor is represented by a complex symmetric fourth rank spinor $\phi_{ABCD}(\bm r,t)$,
\begin{eqnarray}\label{comp1}
G_{\mu\nu\lambda\rho}(\bm r,t)=S_{\mu\nu}^{\;\;AB}S_{\lambda\rho}^{\;\;CD}\phi_{ABCD}(\bm r,t).
\end{eqnarray}
The counterpart of the Riemann-Silberstein vector $\bm F$ is now \cite{mb} a complex symmetric traceless $3\times 3$ matrix $\mathcal{\bm G}_{ij}$ built from the components of $G_{\mu\nu\lambda\rho}$,
\begin{eqnarray}\label{rs1}
{\mathcal G}_{ij}={\mathcal E}_{ij}+\rmi{\mathcal B}_{ij}=G_{0i0j}.
\end{eqnarray}
The six components of $\mathcal{\bm G}$ are connected with the spinorial components as follows:
\begin{eqnarray}\label{phi1}
\fl{\mathcal G}_{11}=\phi_{0000}-2\phi_{0011}+\phi_{1111},\quad
{\mathcal G}_{12}=\rmi\phi_{0000}-\rmi\phi_{1111},\quad{\mathcal G}_{13}=2\phi_{0111}-2\phi_{0001},\nonumber\\
\fl{\mathcal G}_{22}=-\phi_{0000}-2\phi_{0011}-\phi_{1111},\quad
{\mathcal G}_{23}=-2\rmi\phi_{0001}-2\rmi\phi_{0111},\quad
{\mathcal G}_{33}=4\phi_{0011}.
\end{eqnarray}

At first glance, the matrix $\mathcal{\bm G}_{ij}$ does not look like a relativistic object since the formula (\ref{rs1}) seems to distinguish the time direction. However, in analogy with the electromagnetic case, the term {\em relativistic} is justified since $\mathcal{\bm G}_{ij}$ has precisely defined transformation properties under the full Lorentz group. Again, as in the electromagnetic case, relativistic transformation properties of $\mathcal{\bm G}$ follow from the one-to-one correspondence between $\mathcal{\bm G}_{ij}$ and the Riemann tensor; $\mathcal{\bm G}_{ij}$ carries full information about the linearized gravitational field. The matrix ${\hat O}(\Lambda)$ defined in the previous section acting twice on the $\mathcal{\bm G}_{ij}$ matrix produces the correct transformation of the Riemann tensor.

It has been shown by Maartens and Bassett \cite{mb} that the tensor ${\mathcal G}_{ij}$ obeys the evolution equations very similar to the equations for the Riemann-Silberstein vector. They write: ``the analogy with the Maxwell equations is made strikingly apparent in our formalism.'' Hence, the similarity between the electromagnetic field equations and the gravitational field equations can be obtained for the tensorial objects and not only for the objects proposed by Barnett \cite{sb}.

\section{Electromagnetic and gravitational Bessel beams}

Bessel beams are obtained when the generating function is chosen as follows \cite{bb1,bb3}:
\begin{eqnarray}\label{bes}
\chi_m(\rho,\varphi,z,t)=e^{\rmi(k_z z-\omega t+m\varphi)}J_m(\kappa\rho),
\end{eqnarray}
where $\omega=\sqrt{k_z^2+\kappa^2}$. The formula (\ref{sol2}) for the components of $\phi_{AB\dots L}$ has then the following simple form (cf. (\ref{sol2})):
\begin{eqnarray}\label{bes1}
\phi_{00\dots011\dots 1}=(-\rmi)^l \left(\omega+k_z\right)^l\kappa^n\,\chi_{m+n}(\rho,\varphi,z,t).
\end{eqnarray}
Therefore, in the electromagnetic case we obtain:
\begin{eqnarray}\label{bese}
\left[\begin{array}{c}
\phi_{00}\\
\phi_{01}\\
\phi_{11}
\end{array}\right]=
e^{\rmi(k_z z-\omega t+M\varphi)}\left[\begin{array}{c}
-\left(\omega+k_z\right)^2e^{-\rmi\varphi}J_{M-1}(\kappa\rho)\\
-\rmi\left(\omega+k_z\right)\kappa\,J_{M}(\kappa\rho)\\
\kappa^2e^{\rmi\varphi}J_{M+1}(\kappa\rho)
\end{array}\right].
\end{eqnarray}
With the use of (\ref{phi}) we reproduce (up to an overall normalization) the formulas for the electromagnetic Bessel beams obtained before in \cite{bb1,bb3,bbd}. The ``quantum number'' $M=m+1$ is the value (in units of $\hbar$) of the projection of the {\em total angular momentum} on the $z$-axis. It has two parts: the orbital angular momentum $m$ appearing in the formula (\ref{bes}) and the contribution from helicity equal to 1. In order to obtain the wave with the opposite polarization we must reverse in (\ref{bes}) the sign in the exponent.

The extension of these results to the gravitational case is straightforward. The counterpart of the formula (\ref{bese}) reads:
\begin{eqnarray}\label{besg}
\left[\begin{array}{c}
\phi_{0000}\\
\phi_{0001}\\
\phi_{0011}\\
\phi_{0111}\\
\phi_{1111}
\end{array}\right]=
e^{\rmi(k_z z-\omega t+M\varphi)}\left[\begin{array}{c}
\left(\omega+k_z\right)^4e^{-2\rmi\varphi}J_{M-2}(\kappa\rho)\\
\rmi\left(\omega+k_z\right)^3\kappa\,e^{-\rmi\varphi}J_{M-1}(\kappa\rho)\\
-\left(\omega+k_z\right)^2\kappa^2J_{M}(\kappa\rho)\\
-\rmi\left(\omega+k_z\right)e^{\rmi\varphi}J_{M+1}(\kappa\rho)\\
\kappa^4e^{2\rmi\varphi}J_{M+2}(\kappa\rho)
\end{array}\right].
\end{eqnarray}
This solution of the gravitational wave equation, like an electromagnetic Bessel beam, is characterized by three parameters: the frequency $\omega$, the wave vector $k_z$ in the $z$-direction, and the $z$-component of the total angular momentum $M$ equal to $m+2$. The number $m$ refers to orbital angular momentum and 2 is the contribution due to the helicity.

\section{Other beams with OAM}

Bessel beams are simple in their mathematical form but they are not realistic because they carry infinite energy. All monochromatic beams have this fault. Finite energy can be obtained only for non-monochromatic wave packets. Widely used Laguerre-Gauss beams fall into this category. The generating function for such beams with the OAM quantum number $m$ is \cite{bb3}:
\begin{eqnarray}\label{laguerre}
\fl\qquad\chi_{\Omega nm}^\sigma(\rho,\phi,z,t) =\frac{e^{-i\sigma(\Omega(t-z/c)-m\phi)}\rho^m}{a(t_+)^{n+m+1}}
\exp\left(-\frac{\rho^2}{a(t_+)}\right) L_n^m\!\left(\frac{\rho^2}{a(t_+)}\right),
\end{eqnarray}
where $a(t_+)=l^2+i\sigma c^2(t+z/c)/\Omega$, $\sigma$ is the sign of helicity, and $L_n^m$ is the associated Laguerre polynomial. By differentiating this generating function according to (\ref{sol}) we obtain both electromagnetic and gravitational Laguerre-Gauss beams. These beams carry finite energy but they have another unpleasant feature. Even though the energy is predominantly flowing in the positive direction of the $z$-axis, there is an admixture flowing in the opposite direction due to the dependence on $t_+$. There are, however, examples of finite energy analytic solutions carrying orbital angular momentum, called exponential beams, with a unidirectional flow of energy. The generating function $\chi_{mk_z\tau}(\rho,\phi,z,t)$ for one polarization in this case is \cite{bb4}:

\begin{eqnarray}\label{app3}
\chi_{mk_z\tau}(\rho,\phi,z,t) =\frac{\rme^{\rmi(m\phi+k_z z)}\rme^{-\vert k_z\vert s}\rho^m}{s\left(s-c(\tau+\rmi t)\right)^m},
\end{eqnarray}
where $s=\sqrt{\rho^2 - c^2(t-\rmi\tau)^2}$. The functions $\chi_{mk_z\tau}$ describe beams with an exponential fall-off. Indeed, for large $\vert k_z\vert\rho$, at the fixed values of all the remaining variables, these functions decrease exponentially
\begin{eqnarray}\label{expfall}
\chi_{mk_z\tau}(\rho,\phi,z,t)\mathop{\sim}_{\rho\to\infty}
\rme^{\rmi(m\phi+k_z z)}\frac{\rme^{-\vert k_z\vert\rho}}{\rho}.
\end{eqnarray}

The construction of electromagnetic and gravitational waves with the HDBWP method yields also immediately the gravitational counterpart of the Hopfion studied extensively in \cite{tsb,twsb,stwdb}. The generating function $\chi_a(\bi r,t)$ for the Hopfion is:
\begin{eqnarray}\label{syn}
\chi_a(\bi r,t)=\frac{1}{r^2-(t-\rmi a)^2}=\frac{1}{4\pi}\int\!\frac{d^3k}{k}\,
e^{-|a|k}\rme^{\rmi\,\rm{sgn}(a)(\bi k\cdot\bi r-\omega t)}.
\end{eqnarray}
Depending on the sign of $a$, we obtain positive or negative helicity. The generating function (\ref{syn}) has axial symmetry, so that the Hopfion does not carry angular momentum. However, more elaborate knotted solutions studied in \cite{kea} built on the Hopfion will be, in general, endowed with orbital angular momentum.

\section{Particle-wave duality and the electromagnetic-gravitational correspondence}

In order to establish a one-to-one correspondence between the electromagnetic field and the linearized gravitational field we may also use the particle picture: photons and gravitons. To this end we start from the mathematically rigorous theory of the representations of the Poincar\'e group. As has been established by Wigner \cite{wig}, for massless particles there are two (except for spinless particles) unitary inequivalent {\em one-dimensional} representations of the Poincar\'e group, corresponding to two signs of helicity. Wigner already noticed that these representations can be either interpreted as the quantum wave functions of massless particles in momentum space or alternatively as Fourier representations of the classical waves. In the previous sections we used the second interpretation and we treated the waves fields as purely classical objects. However, the first interpretation is also worth considering because it also throws some light on the electromagnetic-gravitational correspondence.

For unitary one-dimensional representations $f_\lambda(\bm k)$ the action of the Poincar\'e group element ${\mathfrak g}$ can only produce a phase factor. As shown by Wigner, this phase factor has the form
\begin{eqnarray}\label{tr}
\begin{CD}
f_\lambda(\bm k)@>{\mathfrak g}>>\rme^{\rmi\lambda H\varphi_{\mathfrak g}(\bm k)}f_\lambda(\bm k),
\end{CD}
\end{eqnarray}
where $H$ is the absolute value of helicity ($H=1$ for photons and $H=2$ for gravitons), $\lambda$ is the sign of helicity, and $\varphi_{\mathfrak g}(\bm k)$ represents the group element ${\mathfrak g}$. In \cite{pwf,bb1,bb} we gave an explicit form of the generators of Poincar\'e transformations acting on the wave functions $f_\lambda(\bm k)$. The phase function $\varphi_{\mathfrak g}(\bm k)$ is universal; the same for all helicities. This universality will enable us to establish a clear-cut correspondence between electromagnetism and gravitation.

Now, we must establish a connection between the wave functions $f_\lambda(\bm k)$ and the local wave fields that live in spacetime. The use of a straightforward Fourier transformation will not do because the wave field $\phi(\bm r,t)$,
\begin{eqnarray}\label{notdo}
\phi_\lambda(\bm r,t)=\int\frac{d^3k}{(2\pi)^{3/2}}\rme^{\rmi{\bm k}\cdot{\bm r}-\rmi\omega t}f^*_\lambda(\bm k),
\end{eqnarray}
is not local \cite{pwf} when the amplitudes $f_\lambda(\bm k)$ are transformed according to (\ref{tr}). Its value at the spacetime point $(\bm r,t)$ after the transformation will depend on its values everywhere. To obtain a local wave field we must insert some additional factors under the integral. This is easily accomplished in the spinorial formalism. It follows from (\ref{sol}) that the general local field (including the contributions from both polarizations) has the form:
\begin{eqnarray}\label{solf}
\fl\qquad\phi_{AB\dots L}(\bm r,t)=\int\frac{d^3k}{(2\pi)^{3/2}}\kappa_A\kappa_B\dots\kappa_L\left(f_+(\bm k)\rme^{\rmi{\bm k}\cdot{\bm r}-\rmi\omega t}
+f^*_-(\bm k)\rme^{-\rmi{\bm k}\cdot{\bm r}+\rmi\omega t}\right),
\end{eqnarray}
where $\kappa_A=g^{\mu}_{\;A\dot{A}}\sigma^{\dot A}k_\mu$ and the wave functions $f_\lambda(\bm k)$ serve as the Fourier components in the expansion of the generating function $\chi(\bm r,t)$ into plane waves,
\begin{eqnarray}\label{genf}
\chi(\bm r,t)=\int\frac{d^3k}{(2\pi)^{3/2}}\left(f_+(\bm k)\rme^{\rmi{\bm k}\cdot{\bm r}-\rmi\omega t}
+f^*_-(\bm k)\rme^{-\rmi{\bm k}\cdot{\bm r}+\rmi\omega t}\right).
\end{eqnarray}
In the second term in these formulas the function $f_-(\bm k)$ enters as the complex conjugate wave function because it is multiplied by the negative energy phase factor.

For electromagnetism and gravitation the formula (\ref{solf}) gives:
\numparts
\begin{eqnarray}\label{soleg}
\fl\qquad{\bm F}(\bm r,t)=\int\frac{d^3k}{(2\pi)^{3/2}}{\bm e}(\bm k)\left(f_+(\bm k)\rme^{\rmi{\bm k}\cdot{\bm r}-\rmi\omega t}
+f^*_-(\bm k)\rme^{-\rmi{\bm k}\cdot{\bm r}+\rmi\omega t}\right),\\
\fl\qquad{\mathcal G}_{ij}(\bm r,t)=\int\frac{d^3k}{(2\pi)^{3/2}}e_i(\bm k)e_j(\bm k)\left(f_+(\bm k)\rme^{\rmi{\bm k}\cdot{\bm r}-\rmi\omega t}
+f^*_-(\bm k)\rme^{-\rmi{\bm k}\cdot{\bm r}+\rmi\omega t}\right),
\end{eqnarray}
\endnumparts
where the polarization vector is built from the products of the spinors $\kappa$,
\begin{eqnarray}\label{pol}
e_i(\bm k)=S_{0i}^{\;\;AB}\kappa_A\kappa_B.
\end{eqnarray}

All wave functions $f_\lambda(\bm k)$ for beams carrying orbital angular momentum have the characteristic dependence $e^{\rmi m\phi}$ on the polar angle $\phi$ in the momentum space \cite{bb3}. This justifies the use of the term {\em quantum number} for $m$.

\section{Conclusions}

The aim of this work was to obtain explicit solutions describing the gravitational waves carrying orbital angular momentum. To this end we have used the fact that every electromagnetic wave and every gravitational wave in the linearized theory is characterized by a single complex generating function: a solution of the scalar d'Alembert equation. Complete solutions are constructed from the second (electromagnetism) or fourth (gravitation) derivatives of the generating function. The use of the same generating function in both cases establishes a one-to-one correspondence between the electromagnetic and gravitational waves.

\section*{Acknowledgments}

This research was financed by the Polish National Science Center Grant No. 2012/07/B/ST1/03347.

\end{document}